\documentclass[aps,prb,reprint,superscriptaddress]{revtex4-1}
\usepackage{amssymb,amsmath,mathptmx}
\usepackage{graphicx}
\usepackage{dcolumn}
\usepackage{bm}
\usepackage{hyperref}
\usepackage{color}
\usepackage[utf8x]{inputenc}
\usepackage{array}
\newcolumntype{C}{>{\centering\arraybackslash}p{1em}}
\newcommand{\kbf}{\mathbf{k}}

\newcommand{\be}{\begin{equation}}
\newcommand{\ee}{\end{equation}}
\newcommand{\bea}{\begin{eqnarray}}
\newcommand{\eea}{\end{eqnarray}}
\begin{document}


\title{Engineering the Floquet spectrum of superconducting multiterminal
  quantum dots}

\author{R\'egis M\'elin}

\affiliation{Univ. Grenoble-Alpes, CNRS, Grenoble INP\thanks{Institute
    of Engineering Univ. Grenoble Alpes}, Institut NEEL, 38000
  Grenoble, France}

\author{Romain Danneau}
\affiliation{Institute of Nanotechnology, Karlsruhe Institute of Technology, D-76021 Karlsruhe, Germany}

\author{Kang Yang}

\affiliation{Laboratoire de Physique Théorique et Hautes Energies, Sorbonne Universit\'e and CNRS UMR 7589, 4 place Jussieu, 75252 Paris Cedex 05, France}

\affiliation{Laboratoire de Physique des Solides, CNRS UMR 8502,
  Univ. Paris-Sud, Universit\'e Paris-Saclay F-91405 Orsay Cedex, France}
\author{Jean-Guy Caputo}

\affiliation{Laboratoire de Math\'ematiques, INSA de Rouen, Avenue de
  l'Universit\'e, F-76801 Saint-Etienne du Rouvray, France}

\author{Benoît Douçot}
\affiliation{Laboratoire de Physique Théorique et Hautes Energies, Sorbonne Universit\'e and CNRS UMR 7589,  4 place Jussieu, 75252 Paris Cedex 05, France}

\date{\today}
\begin{abstract}

Here we present a theoretical investigation of the Floquet spectrum in
multiterminal quantum dot Josephson junctions biased with commensurate
voltages.  We first draw an analogy between the electronic band theory
and superconductivity which enlightens the time-periodic dynamics of
the Andreev bound states. We then show that the equivalent of the
Wannier-Stark ladders observed in semiconducting superlattices
\textit{via} photocurrent measurements, appears as specific peaks in
the finite frequency current fluctuations of superconducting
multiterminal quantum dots.  In order to probe the
Floquet-Wannier-Stark ladder spectra, we have developed an analytical
model relying on the sharpness of the resonances. The charge-charge
correlation function is obtained as a factorized form of the Floquet
wave-function on the dot and the superconducting reservoir
populations. We confirm these findings by Keldysh Green's function
calculations, in particular regarding the voltage and frequency
dependence of the resonance peaks in the current-current
correlations. Our results open up a road-map to quantum correlations
and coherence in the Floquet dynamics of superconducting devices.

\end{abstract}

\maketitle

\section{Introduction}

Since the early 1960’s, the Josephson effect has attracted continuous
interest and its development over the years has led to major
applications in quantum information and technologies. It occurs when
two superconductors connect a nonsuperconducting material and its
physical mechanism can be described {\it via} the notion of Andreev
bound states (ABS). These phase-sensitive midgap doublets are produced
by proximity effect. The ABS microscopically originate from the
formation of entangled electron–hole pairs in the normal conductor and
can be seen as two-level systems \cite{Zazunov}.  Many physical
properties of the Josephson junctions, such as the value of the
Josephson current, depend on the energy-phase relation of the ABS at
zero bias voltage and finite phase drop across the junction
\cite{Andreev,deGennes,SaintJames,Kulik}.

The growing interest in quantum information has boosted investigations
on the zero-energy states in proximitized superconducting structures
\cite{Kitaev}. {ABS physics is pivotal in the interpretation of the
  experimental evidence for Majorana bound states in nanowires
  \cite{Mourik,Huang,Badiane1,Houzet1,Badiane2}. Moreover, new states
  of matter} based on ABS have been predicted in conventional
superconductor multi-terminal devices, offering the possibility to
engineer artificial topological materials featuring Weyl singularities
\cite{vanHeck,Padurariu,Riwar,Strambini}. We note that recent works
aiming at probing these topological systems have been reported
\cite{Draelos,Pankratova}.  Experiments on multiterminal
superconducting junctions have been already performed
\cite{Pfeffer,Cohen,Note1}, highlighting multiple Andreev {reflections
  (MAR)} involving more than two leads \cite{Houzet2,Melin1,Nowak} as
well as correlations between Cooper pairs
\cite{Cuevas1,Freyn1,Jonckheere,Rech,Melin2}. In addition, ABS in the
static regime have also been proposed to create triplet correlations
using {ferromagnetic wires} in multiterminal configurations
\cite{Mai}, to study the effect of spin-orbit interactions in 1D
systems coupled to superconducting leads \cite{Murani} and to simulate
Andreev molecules using two Josephson junctions in series
\cite{Pillet1}.

ABS have been experimentally studied by tunnel or microwave
spectroscopy
\cite{Pillet2,Dirks,Bretheau1,Bretheau2,Schindele,Olivares,Janvier,Gramich1,Bretheau3,Gramich2,Dassonneville,Tosi}.
Theoretically, one can distinguish between two different regimes: the
first one refers to a static phase configuration, \textit{i.e.} when a
Josephson junction is driven by a time-independent magnetic flux
within a loop, with all parts of the circuit at the same chemical
potential. The second one corresponds to a dynamical control of the
superconducting phases, for instance when the system is set
out-of-equilibrium \textit{via} voltage biasing. The latter is a
time-periodic problem and therefore described by the Floquet theory
where time periodicity plays the role of spatial periodicity for
electrons in a solid. Recently \cite{Melin3}, some of the authors have
shown that, set out-of-equilibrium, the ABS appear as periodic
resonances equivalent to the Wannier-Stark ladders
predicted~\cite{Wannier} for a solid under electric field, and
observed in semiconducting superlattices from photocurrent
experiments~\cite{Mendez1,Mendez2}. In these superconducting systems,
the time-periodicity of the ABS dynamics implies the emergence of a
spectrum made of two sets of energy levels, namely the
Floquet-Wannier-Stark (FWS) ladders.

\begin{figure}[htb]
  \includegraphics[width=.8\columnwidth]{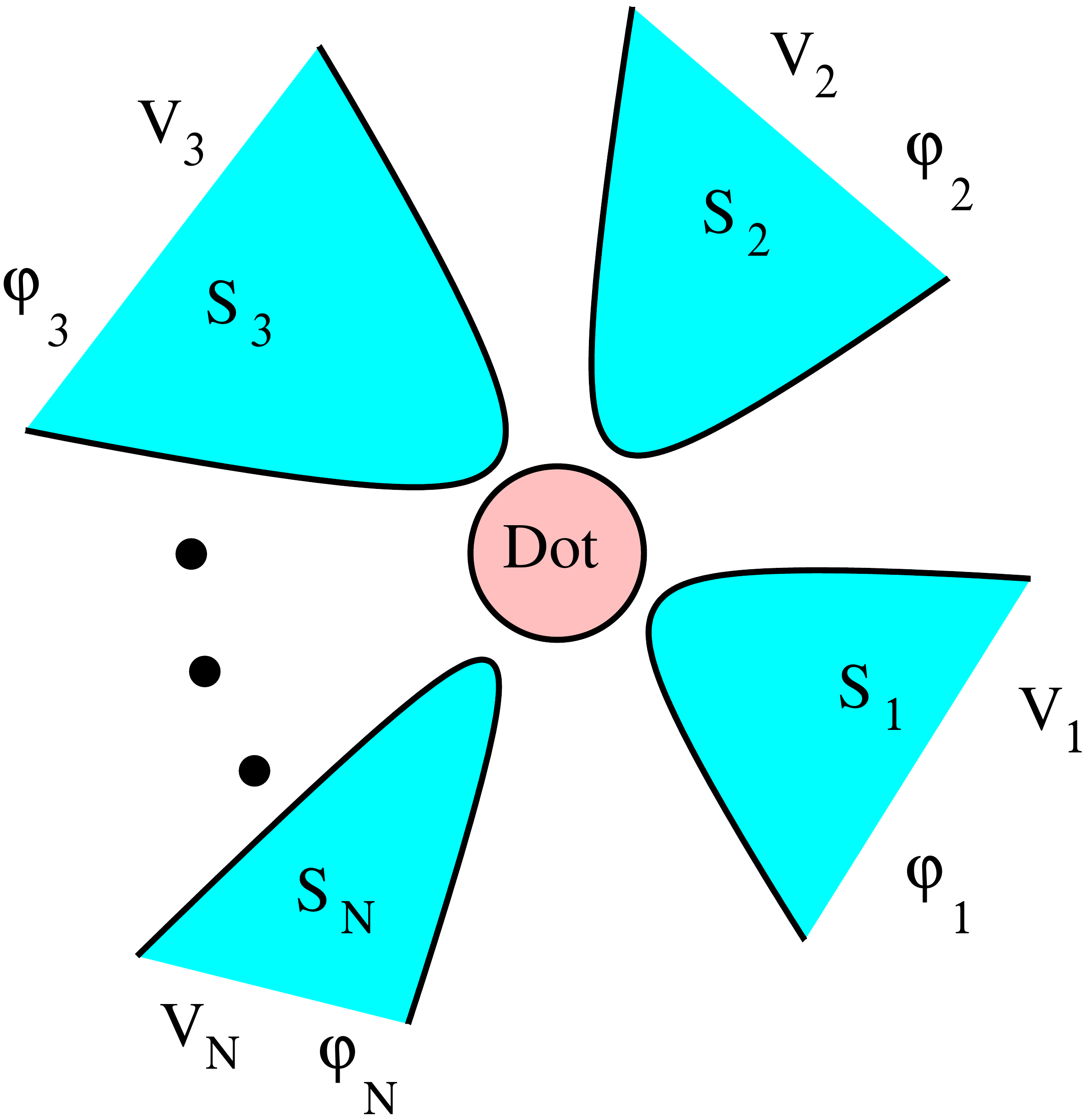}
  \caption{A $N$-terminal superconducting-quantum dot biased at
    voltages $V_1,\, ...,\,V_N$. The superconducting phase of lead
    $S_n$ evolves according to $\varphi_n(t)=\varphi_n+2e V_n
    t/\hbar$, where $\varphi_1,\,...,\,\varphi_N$ are the phases at
    the origin of time $t=0$. Commensurate ratio between the $V_n$ is
    assumed in the paper. The resonant quantum dot hosts a single
    spin-degenerate level at zero energy.
      \label{fig:schema-N-terminaux}
    }
\end{figure}

In this paper, we study the Floquet spectrum of a superconducting
multiterminal quantum dot (QD) by means of analytical and numerical
calculations.  We show that in this configuration (see
Fig.~\ref{fig:schema-N-terminaux}) the FWS ladders can be revealed by
finite frequency noise spectroscopy, with sharp peaks at the
transitions between pairs of FWS resonances. Our approach involves an
analytical calculation of the charge-charge correlation function which
shows that the peak frequencies obtained in the noise match the energy
spacing between arbitrary levels in the Floquet spectra.  The
analytical model based on a sharp resonance approximation is used to
label the noise spectra obtained numerically from microscopic Keldysh
Green's function calculations. Our work enables further investigations
of the coherent qu-bit-like dynamics of two FWS ladders.

This paper is organized as follows. In section~\ref{section_parallel},
we expose an analogy between Wannier-Stark ladders in band theory and
in multiterminal hybrid superconducting systems. The Hamiltonian is
provided in section~\ref{sec:Hamiltonians}. The results on the
connection between the FWS resonance spectra and finite frequency
cross-correlations are presented in section~\ref{sec:results}. Summary
and perspectives are provided in section~\ref{Conclusion}.

\begin{figure}[htb]
  \includegraphics[width=\columnwidth]{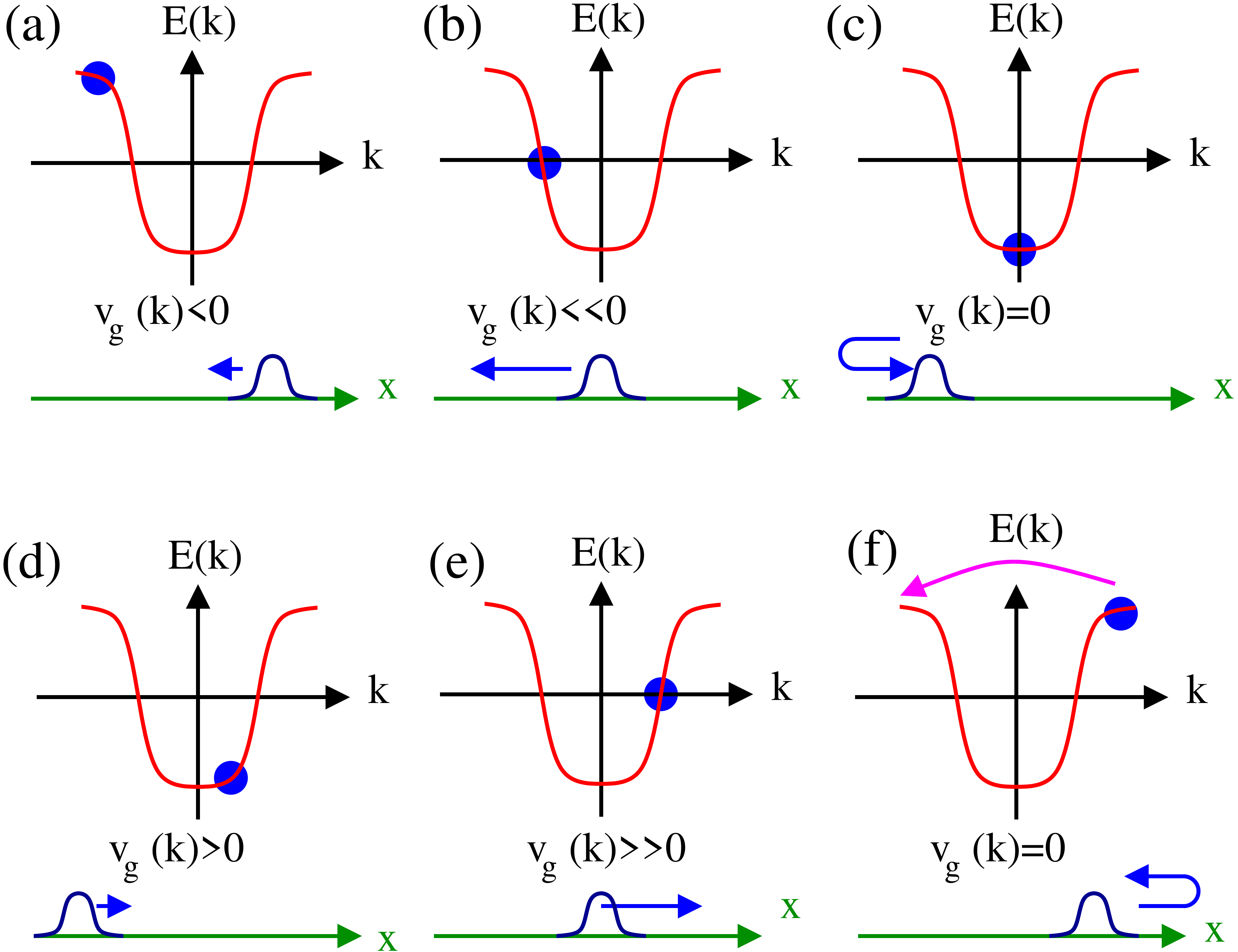}
    \caption{{\it Bloch oscillations:} Evolution of wave-vector $k$
      and real space coordinate $x$ in Bloch oscillations. Wave-vector
      $k$ increases linearly in time according to $dk/dt = -eE/\hbar$,
      with $E$ the electric field. The group velocity $v_g (t) =
      dE/dk$ oscillates as a function of time $t$, yielding
      oscillations of wave-packets in real space, with frequency
      proportional to the electric field $E$. Panels (a)-(f) cover one
      period of oscillations.  Due to the periodicity of the lattice
      potential in real space, the Fourier point $\pi/a’_0$ in panel
      (f) is identified to $-\pi/a’_0$ in panel (a).
      \label{fig:Bloch-oscillations}
    }
\end{figure}

\section{Parallel between band theory and superconductivity}
\label{section_parallel}

{Following the seminal works of Anderson
  \cite{Anderson1,Anderson2}, a classical parallel is known between
  two distinct fields of condensed matter physics, \textit{i.e.} band
  theory and superconductivity.  We go beyond by implementing this
  analogy for time-periodic Floquet Hamiltonians.}

\paragraph*{Bloch oscillations in periodic crystals:}

A simple cubic lattice crystal is parameterized by the spacing $a_0$
between nearest neighboring sites. Electrons are localized
independently on each atom if $a_0$ is much larger than the size of
the atomic electronic clouds. As $a_0$ diminishes, the ground state
degeneracy for the single electron Hamiltonian is gradually lifted
upon increasing tunnel coupling between neighboring electronic
clouds. A ``band'' of energy with continuous spectrum is then formed
in the thermodynamic limit. This corresponds to band theory with Bloch
wave-functions \cite{Bloch}. Eigenstates are extended plane waves
multiplied by a function which is periodic in $a_0$.

In the presence of an additional static and uniform electric field
$E$, Zener \cite{Zener} showed theoretically that an electron in a
crystal oscillates periodically in space, and that electromagnetic
radiation is emitted at the corresponding frequency (see
Fig.~\ref{fig:Bloch-oscillations}). However, the so-called Bloch
oscillations have never been demonstrated experimentally for bulk
materials (neither bulk metals nor bulk semiconductors). In this case,
the inelastic scattering time is much shorter than the
{delay $\Delta t=h/eEa_{0}$ for crossing the Brillouin
  zone under action of} electric field.  However, $\Delta t$ is
strongly reduced in artificial semiconducting superlattices, in which
the potential can be modulated with period $a'_0$, much larger than
the lattice parameter $a_0$ of a bulk semiconductor. The
semiconducting superlattice Brillouin zone (having a size $\sim
1/a'_0$) is thus strongly reduced compared to that of the
corresponding bulk semiconductor (having a size $\sim 1/a_0$). The
delay for crossing the semiconducting superlattice Brillouin zone
under action of the electric field can advantageously be much smaller
than the inelastic scattering time \cite{Esaki}. Many cycles of Bloch
oscillations are then possible on time scale much shorter than the
inelastic scattering time, making possible the observation of Bloch
oscillation-related effects. Bloch oscillations are the time-dependent
counterpart of the Wannier-Stark ladder spectrum mentioned in the
Introduction. Indeed, the spectral gap $eEa_{0}$ between two
consecutive energy levels in such ladder is equal to $h \nu_{B}$,
where $\nu_{B}=1/\Delta t$ is the frequency associated to Bloch
oscillations. This has been observed in ultra-cold atoms
\cite{Wilkinson,Dahan,Geiger}. We notice the properties of Bloch
oscillations have already been used in Coulomb blockaded Josephson
junction circuits for low noise amplification \cite{Delahaye,Sarkar}.

\paragraph*{Parallel between band theory and superconductivity:}

The use of conventional superconductors (such as aluminum) in
electronic devices based on the Josephson effect is at the heart of
the developments on quantum circuits and quantum technologies. One of
the reasons why superconductivity remains forefront in both fundamental
and applied physics research for more than a century is reflected in
the existence of broken gauge symmetry. The BCS microscopic theory
describes superconductivity as an effect of electron-phonon coupling
yielding formation of bound states of electron pairs with
opposite-spins, so-called ``Cooper pairs'', which condense into a
collective ground state. Anderson implemented his theory of gauge
invariance which successfully accounts for the Meissner effect on the
basis of the so-called Higgs mechanism
\cite{Anderson1,Anderson2}. Even if the phase of a single
superconductor is not measurable, phase differences are
gauge-invariant (and thus measurable) quantities. Consequently, the
Josephson effect occurs as a dissipationless current flowing through a
weak link \cite{Likharev} connected by two phase biased
superconductors.

It turns out that band theory and superconductivity share deep common
features.  Based on the concept of phase rigidity \cite{Anderson1},
Anderson has described the wave function associated to the Josephson
effect as coherent superposition between states with different numbers
of pairs within the two superconducting leads \cite{Anderson2}. In
solid state physics, wave-vectors in the Brillouin zone are analogous
to superconducting phases between $0$ and $2\pi$ while the
``position'' basis corresponds to the ``number of transmitted pairs''
basis in superconductivity. The Wannier-Stark ladders emerge in both
cases as natural consequence of this analogy.  The parallel between
band theory and superconductivity is described in Table~\ref{tab}.

\begin{table}[!htb]
  {
   \begin{tabular}{|*{2}{>{\centering\arraybackslash}p{.5\columnwidth}|}}
      \hline
      \setlength\arraycolsep{10pt}
      {\bf Band Theory} & {\bf Superconductivity}\\
      \hline
      \hline
      Wave-vectors & Superconducting phases\\
      \hline
      Position $x_n$ in & Number $N$ of transmitted\\
      real space & Cooper pairs\\
      \hline
      $x_{n}/a_{0}$ integer equivalent to & $N$ integer equivalent to\\
      $2\pi/a_{0}$-periodic wave-vector $k$ &
      $2\pi$-periodic phase $\varphi$\\   
      \hline
      Plane waves in Bloch theory &
      States with fixed superconducting\\
      $|k\rangle=\sum_x\exp(ikx) |x\rangle$  &
      phase $|\varphi\rangle=\sum_N\exp(iN\varphi)|N\rangle$\\
      \hline
      Hopping between neighboring&
      Transferring pairs between leads\\
      tight-binding sites & by Andreev reflection\\
      \hline
      Electric field & Josephson relation \\
      $dk/dt=-e E$ &
      $d \varphi_n/dt=2eV_n/\hbar$\\
      \hline
          {\bf Wannier-Stark ladders} & {\bf Floquet-Wannier-Stark ladders} \\
          \hline
          \hline
    \end{tabular}
  }
  \caption{\label{tab} Analogy between band theory and
    superconductivity.}
\end{table}

\begin{figure*}[htb]
    \includegraphics[width=.99\textwidth]{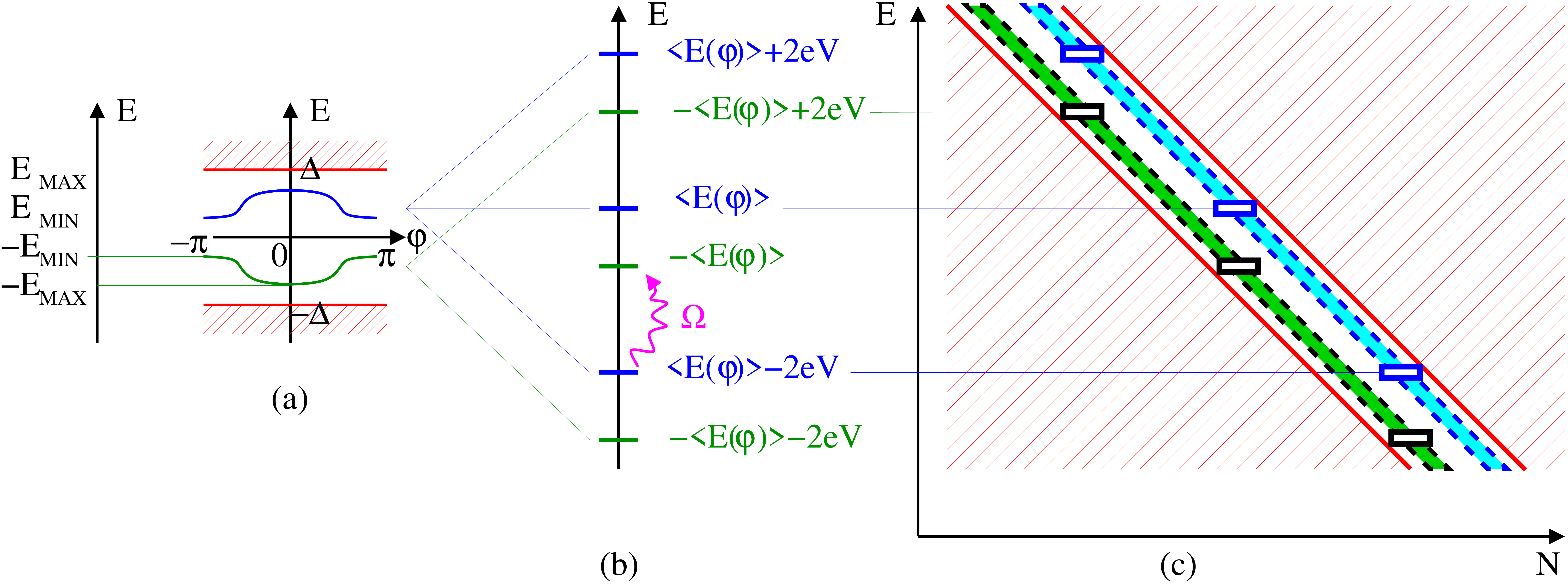}
    \caption{{\it FWS ladders:} (a) Energies $\pm E(\varphi)$ of the
      two ABS as a function of superconducting phase variable
      $\varphi$ in absence of bias voltage ($V=0$). The red dashed
      regions correspond to the quasiparticle continua. (b) The
      corresponding FWS ladders in presence of finite bias voltage
      $V$.  Notation $\langle E(\varphi)\rangle$ is used for the
      average of $E(\varphi)$ over the superconducting phase variable
      $\varphi$. (c) Energy \textit{vs.} number $N$ of transmitted
      Cooper pairs displaying the tilted band picture of FWS ladder
      localization.
      \label{fig:ladders}
      }
\end{figure*}

Having emphasized this analogy, it is worth pointing out a distinction
between momentum $k$ in band theory and the phase difference $\varphi$
in superconductivity. Whereas $k$ has to be regarded as a {\em good
  quantum number}, $\varphi$ is a {\em classical parameter} in the BCS
mean-field Hamiltonian.  However, one can circumvent this issue by
promoting $\varphi$ to a genuine quantum degree of freedom,
canonically conjugate to the number of transmitted Cooper pairs $N$
(see Table~\ref{tab}).

{It turns out that superconductivity is not required
  for producing dynamical Wannier-Stark ladders \cite{Hino2008}. In
  this work\cite{Hino2008}, the authors study theoretically the
  response of a semiconducting superlattice to a periodic train of
  pulses of the electric field. However, superconductivity offers the
  unique opportunity to explore Floquet physics with purely
  dc-voltage biasing.}

\paragraph*{How to detect FWS ladders?}

In a two-terminal Josephson junction biased with voltage $V$, the
superconducting phase $\varphi(t)$ winds in time $t$ according to the
Josephson relation
\begin{equation}
  \label{eq:relation-Josephson}
  \varphi(t) = \varphi+ 2eVt/\hbar
  .
\end{equation}
Then, the two equilibrium ABS [see Fig.~\ref{fig:ladders} (a)] give
rise to two alternating FWS ladders [see Fig.~\ref{fig:ladders} (b)]
\cite{Melin3}, which are the counterparts of the Wannier-Stark ladders
\cite{Wannier} observed experimentally in semiconducting superlattices
\cite{Mendez1,Mendez2}.

This raises the natural {question of} how to demonstrate
experimentally the presence of FWS ladders, and to extract their
precise location in energy. The two ladders are indeed at energies
(see Fig.~\ref{fig:ladders})
\begin{equation}
  \label{eq:energy-shift}
  {E}_{q,\pm}=E_{\pm}+qeV/\hbar
\end{equation}
($q$ being an even integer for our two-terminal junction, and
with any parity for three terminals). The simple relation $E_+=-E_-$
for the energy shifts $E_\pm$ is valid in general. Far from crossing
in the real part of FWS resonances, we have in addition $E_+\simeq
\langle E_{ABs}\rangle$, where $\langle E_{ABS} \rangle$ is the
average of the equilibrium ABS energy over the fast superconducting
phase variable\cite{Melin3}.

{Microwave radiation} can excite transitions between two arbitrary
rungs $E_{q_1,\epsilon_1}$ or $E_{q_2,\epsilon2}$ [see
  Eq.~(\ref{eq:energy-shift}), with $q$ replaced by $q_1$ or $q_2$,
  and $\epsilon_{1,2}=\pm$], on the condition of resonance
$\Omega=E_{q_2,\epsilon_2} - E_{q_1,\epsilon_1}$ between the rf-field
frequency $\Omega$ and the energy difference
$E_{q_2,\epsilon_2}-E_{q_1,\epsilon_1}$ [see
  Eq.~(\ref{eq:energy-shift})], as illustrated by the drawing on
Fig.~\ref{fig:ladders} (b).  It is also possible to perform such
spectroscopy by measuring current correlations at finite frequency, as
it will be further explained {in section~\ref{sec:results}. There, we
  will show that, as a function of the measurement frequency $\Omega$,
  the finite frequency current cross-correlations $S_{a,b}(\Omega)$
  exhibit peaks at $\Omega=\Delta E_{+,p},\,\Delta E_{0,p}, \, \Delta
  E_{-,p}$, with
\begin{eqnarray}
  \label{eq:E+}
  \Delta E_{+,p}&=& E_{q_2,+}-E_{q_1,-} =E_+ -E_- + peV/\hbar\\
  \label{eq:DeltaE0p}
  \Delta E_{0,p}&=&E_{q_2,\epsilon}-E_{q_1,\epsilon}=peV/\hbar\\
  \Delta E_{-,p}&=&E_{q_2,-}-E_{q_1,+}=E_--E_+ +peV/\hbar
  \label{eq:E-}
  ,
\end{eqnarray}
where $\epsilon=\pm$, and $p=q_2-q_1$ is an even integer in the case
of two terminals, but takes any parity for three-terminal systems with
commensurate dc-voltage biasing, such as opposite voltages $V_{a}=-V$,
$V_{b}=V$ and $V_{c}=0$ in the quartet configuration \cite{Freyn1}.
Thus, $\Delta E_{+,p}$ and $\Delta E_{-,p}$ encode inter-ladder
transitions, while $\Delta E_{0,p}$ corresponds to intra-ladder
transitions.

\section{Hamiltonians}
\label{sec:Hamiltonians}

We consider in the paper a QD coupled to $N$ superconducting
reservoirs (see Fig.~\ref{fig:schema-N-terminaux}). The reservoirs are
assumed to be biased at dc voltages $V_{i}$ ($1 \leq i \leq N$)
(chosen to be commensurate). Therefore, we write $V_{i}=s_{i}V$ where
$s_{i}$ is an integer. For example, in the quartet configuration
\cite{Freyn1}, we have $N=3$, and $s_{i}\in \{0,1,-1\}$. It is easy to
specialize to the two-terminal case, simply by setting to zero the
tunneling coupling $J_{c}$ to the reservoir such that $s_{c}=0$. Then
the dc voltage drop between the two remaining superconducting
reservoirs is equal to $2V$.

The Hamiltonian can be written as
\begin{equation}
{\cal H}(t)={\cal H}_0+{\cal H}_J(t),\label{eqhmfull}
\end{equation}
where ${\cal H}_0$ is an usual BCS Hamiltonian for the superconducting
reservoirs and ${\cal H}_J(t)$ describes the tunneling processes
between these reservoirs and the QD. Specifically:
\begin{widetext}
\begin{equation}
    {\cal H}_0 = \sum_{j=1}^{N} \sum_{\sigma} \int
    \frac{d^{D}\kbf}{(2\pi)^D}
    \left(\epsilon(j,\kbf)c^\dagger_{\sigma}(j,\kbf)c_{\sigma}(j,\kbf)
    + \Delta_j
    c^\dagger_{\uparrow}(j,\kbf)c^\dagger_{\downarrow}(j,-\kbf)
    +\Delta_{j}^{*}c_{\downarrow}(j,-\kbf)c_{\uparrow}(j,\kbf) \right)
\end{equation}
and
\begin{equation}
  \label{eq:HJ}
  {\cal H}_J = \sum_{j=1}^{N} J_{j} \sum_{\sigma} \int
  \frac{d^{D}\kbf}{(2\pi)^D}
  \left(e^{-is_{j}\omega_{0}t}c^\dagger_{\sigma}(j,\kbf)d_{\sigma}+
  e^{is_{j}\omega_{0}t}d^{\dagger}_{\sigma}c_{\sigma}(j,\kbf)\right)
  .
\end{equation}
\end{widetext}
Here $c^\dagger_{\sigma}(j,\kbf)$ and $c_{\sigma}(j,\kbf)$ are
creation and annihilation operators for an electron on reservoir $j$
with momentum $\kbf$ and spin $\sigma$ along the quantization
axis. Corresponding operators on the dot are denoted by
$d^{\dagger}_{\sigma}$ and $d_{\sigma}$. The dimension $D$ of the
reservoirs is left undetermined, since its actual value is not
crucial. The basic frequency $\omega_{0}$ is associated to single
electron tunneling processes, and it is equal to
$\omega_{0}=eV/\hbar$. Note that $\omega_{0}=\omega_{J}/2$, where
$\omega_{J}$ is the Josephson frequency associated to $V$.

\section{Results}
\label{sec:results}

The ``sharp resonance'' approximation is first introduced at
moderately low voltage in section~\ref{section_reduction}. In the next
section~\ref{sub_charge_noise}, an analytical expression of the finite
frequency charge-charge correlations is obtained with this sharp
resonance approximation. Numerical results for the finite frequency
current-current correlations are presented next in
section~\ref{sub_current_noise}.

\subsection{Sketch of the sharp resonance approximation}
\label{section_reduction}

The goal of this section is to develop an approximation scheme to
evaluate analytically the resolvent ${\cal R}(E)$ at energy $E$,
defined as
\begin{equation}
  \label{eq:R}
        {\cal R}(E)=\left(E-{\cal H}\right)^{-1}
        ,
\end{equation}
where the Hamiltonian ${\cal H}$ is given in
section~\ref{sec:Hamiltonians} [see
  Eqs.~(\ref{eqhmfull})-(\ref{eq:HJ})]. The resulting compact form of
${\cal R}(E)$ resulting from the sharp resonance approximation will be
used in section~\ref{sub_charge_noise} to provide an analytical
expression for the finite frequency charge-charge correlation
function.

Break junction experiments realize superconducting weak links with
only a few conduction channels \cite{Averin,Cuevas2,Scheer}. In a
single-channel weak link \cite{Beenakker}, the ABS are at energies
$\pm \Delta$ if the phase bias $\varphi=0$ vanishes, whatever contact
transparency (with $\Delta$ the superconducting gap). Once biased at
voltage $V$, the superconducting phase difference evolves in time
according to Eq.~(\ref{eq:relation-Josephson}). Consequently, the ABS
touch periodically the gap edge singularities at energies $\pm \Delta$
in the presence of bias voltage: the narrow equilibrium ABS acquire
large width at nonequilibrium. Indeed, numerical calculations
\cite{Cuevas2} of the first harmonics of the current in a
superconducting weak link show smooth energy dependence, without sharp
resonances (see Figs.~3 and 4 in Ref.~\onlinecite{Cuevas2}).

The situation is quite different in superconducting-QD where the
equilibrium ABS stay away from the gap edge singularities in the full
range of the superconducting phase difference $\varphi$, even if
$\varphi=0$. Indeed, the ABS energies at $\varphi=0$ are of order
$\approx \Gamma$, with $\Gamma=J^2/W$. (In this expression, $J$ is the
hopping matrix element between the QD and the superconductor, and $W$
the band-width of the superconducting leads. In experiments
\cite{Cohen}, the parameter $\Gamma$ is usually a fraction of the
superconducting gap $\Delta$). In a voltage-biased superconducting-QD,
the equilibrium ABS are separated from the gap edge singularities by
the finite energy difference $\approx \Delta-\Gamma$. Still, at finite
$V$, the FWS resonances remain coupled by MAR to the semi-infinite
quasiparticle continua. The resulting line-width broadening $\gamma
\approx \Delta \exp(-c\Delta/eV)$ is exponentially small in the ratio
between the gap $\Delta$ and the voltage energy $eV$ \cite{Melin2}
[with $c$ a constant of order unity]. Since $\gamma$ drops rapidly to
zero as $\Delta/eV$ is increased above unity, another mechanism of
relaxation has to be advocated at inverse voltages larger than
$3 \alt \Delta/eV$, such as the coupling to phonons
\cite{Melin3}.

Given the exponential dependence on inverse voltage of the MAR
line-width broadening $\gamma$, we discuss now (within the
wave-function approach introduced in section~I of the Supplemental
Material) an approximation relying on the sharpness of the FWS
resonances for $3 \alt \Delta/eV$ (see sections~I, II and III in
the Supplemental Material \cite{supplemental} for details). A central
role in these calculations is played by ${\cal R}(E)$ in
Eq.~(\ref{eq:R}) which can be factorized according to \be
\mathcal{R}(\tilde{E}+p\omega_{0})_{m,n}\simeq
\sum_{\alpha=\pm}\frac{\Psi_{m+p}(E_{\alpha})\otimes\Phi_{n+p}(E_{\alpha})}
    {\tilde{E}-E_{\alpha}+i\Gamma_{\alpha}}
    \label{main_approximation_for_R},
    \ee where $p$ is an integer. Eq.~(\ref{main_approximation_for_R})
    above is identical to Eq.~(18) in section~II of the Supplemental
    Material \cite{supplemental}. Further details on its demonstration
    can be found in Appendix~A of this Supplemental Material
    \cite{supplemental}.  The notation
    $\{\Psi_{m}(E)\}_{m\in\mathbb{Z}}$ stands for the two-component
    right zero-eigenvector of the Floquet equations. The notation
    $\{\Phi_{n}(E)\}_{n\in\mathbb{Z}}$ is used for the corresponding
    left eigenvector of the transposed equations. The notation
    $\tilde{E}+p\omega_0$ stands for the energy, where $\omega_0=eV$
    and $E_\alpha$ is the energy shift of the FWS ladder $\alpha=\pm$
    [see Eq.~(\ref{eq:energy-shift})]. The notation $\Gamma_\alpha$ is
    used for the corresponding line-width broadening.

    \subsection{Finite frequency charge-charge correlation function
      in the sharp resonance approximation}
\label{sub_charge_noise}
Now, we demonstrate Eqs.~(\ref{eq:E+})-(\ref{eq:E-}) in the sharp
resonance approximation, on the example of the charge-charge
correlation function \bea \label{eq:C} C(t,t') & = &
\sum_{\sigma,\sigma'}\langle
d^{\dagger}_{\sigma}(t)d_{\sigma}(t)d^{\dagger}_{\sigma'}(t')d_{\sigma'}(t')\rangle
\\ & & \mbox{} - \left(\sum_{\sigma}\langle
d^{\dagger}_{\sigma}(t)d_{\sigma}(t)\rangle\right)
\left(\sum_{\sigma'}\langle
d^{\dagger}_{\sigma'}(t')d_{\sigma'}(t')\rangle\right) \nonumber, \eea
where $d_\sigma$ and $d^{\dagger}_\sigma$ are defined in the Appendix.

The sharp resonance approximation discussed above in the preceding
section~\ref{section_reduction} leads to simple expressions for i) the
charge-charge correlation function given by Eq.~(\ref{eq:C}) (see
below), ii) the dot propagators (see section~II in the Supplemental
Material \cite{supplemental}), iii) the charge on the dot
(section~III) and iv) the dc-currents (see section~IV, again in the
Supplemental Material\cite{supplemental}). Our analytical calculations
can be also extended straightforwardly to the finite frequency current
cross-correlations (the expression of which is not given here),
however with more tedious formula.

Eq.~(\ref{eq:C}) is first Fourier transformed from times $t,\,t'$ to
frequencies $\omega,\,\omega'$. The resulting $C(\omega,\omega')$ has
nonvanishingly small elements if $\omega-\omega'$ is an integer
multiple of $\omega_{0}$.  Here, we limit the analysis to the
``diagonal'' time-translational invariant part of the finite frequency
charge correlation function $C_{d}(\Omega) \equiv C(\Omega,\Omega)$,
which takes the following form in the sharp resonance approximation
[similar to Eqs.~(24)-(25) in the Supplemental Material
  \cite{supplemental}]:
\begin{widetext}
  \be
  \label{eq:Cd}
C_{d}(\Omega)=4\sum_{i,j=1}^{N}\sum_{\alpha=\pm}\sum_{\beta=\pm}\sum_{p}^{(i,j)}\frac{S_{i,\alpha}S_{j,\beta}(\Gamma_{\alpha}+\Gamma_{\beta})}
{(\Omega-{E}_{\alpha}-{E}_{\beta}-p\omega_{0})^{2}+(\Gamma_{\alpha}+\Gamma_{\beta})^{2}}
\sum_{(m,m')\in\mathbb{Z}^{2}}^{(i,j)} F_{\alpha,\beta}(m,m',m-p,m'-p)
,
\ee
with
\bea
\label{def_S_i_alpha}
S_{i,\alpha} & = & \frac{\mathcal{A}_{D}J_{i}^{2}}{2(2\pi)^{D-1}\Gamma_{\alpha}}\sum_{p\in\mathbb{Z}}\sum_{\tau}
\nu(i,\alpha,p,\tau)(k(i,\alpha,p,\tau))^{D-1} \theta({E}_{\alpha}+p\omega_{0}-|\Delta_{i}|) \\ \nonumber
& & |\Phi({E}_{\alpha})_{-s_{i}+p,u}e^{i\varphi_{i}/2}x(i,k(i,\alpha,p,\tau))-
\Phi({E}_{\alpha})_{s_{i}+p,v}e^{-i\varphi_{i}/2}y(i,k(i,\alpha,p,\tau))|^{2}
,
\eea
and
\bea
\label{eq:F}
F_{\alpha,\beta}(m,m',n,n') &=&
\Psi_{m,u}({E}_{\alpha})\Psi^{*}_{m',u}({E}_{\alpha})
\Psi_{n,v}({E}_{\beta})\Psi^{*}_{n',v}({E}_{\beta}) +
\Psi_{m,v}({E}_{\alpha})\Psi^{*}_{m',u}({E}_{\alpha})
\Psi_{n,u}({E}_{\beta})\Psi^{*}_{n',v}({E}_{\beta})
,
\eea
\end{widetext}
where the expression of $S_{i,\alpha}$ in Eq.~(\ref{def_S_i_alpha})
above coincides with Eq.~(25) in the Supplemental Material
\cite{supplemental}. The right and left ``Floquet wave-functions'' are
denoted by $\Psi$ and $\Phi$ [see Eq.~(\ref{eq:R}) above].  The
Heaviside step-function is denoted by $\theta$ in
Eq.~(\ref{def_S_i_alpha}). The superconducting leads have dimension
$D$ in Eqs.~(\ref{eq:Cd})-(\ref{eq:F}). The notation ${\cal A}_D$
stands for the $D$-dimensional sphere area (${\cal A}_1=2$, ${\cal
  A}_2=2\pi$, ${\cal A}_3=4\pi$). The integers $s_i$ are used for
characterizing commensurate voltage biasing: The voltage $V_i$ on
superconducting lead $S_i$ is given by $V_{i}=s_{i}V$ [see also the
  Appendix]. The variable $J_i$ is the tunneling amplitude between the
quantum dot and the superconducting lead $S_i$ phase $\varphi_i$. The
integer $p$ in Eq.~(\ref{eq:Cd}) has the same parity as $s_{i}+s_{j}$,
and $m$ and $m'$ have the same parity as $s_{i}$.

The BCS quasiparticle dispersion relation in lead $j$ is given by \be
E(j,k)\equiv
\sqrt{\epsilon(j,k)^{2}+|\Delta_{j}|^{2}}={E}_{\alpha}+p\omega_{0}
\label{implicit_k}
,\ee where $\epsilon(j,k)$ is the kinetic
energy. Eq.~(\ref{implicit_k}) has two solutions labeled by
$\tau\in\{>,<\}$. The density of states $\nu(j,\alpha,p,\tau)$ is
defined as follows: \be \nu(j,\alpha,p,\tau)=\left(\frac{dE(j,k)}{dk}
(k=k(j,\alpha,p,\tau))\right)^{-1} .  \ee The notations $x$ and $y$ in
Eq.~(\ref{def_S_i_alpha}) stand for the BCS coherence factors
\begin{eqnarray}
  x(j,\kbf)&=&\sqrt{\frac{1}{2}
    \left(1+\frac{\epsilon(j,\kbf)}{E(j,\kbf)}\right)}\\
  y(j,\kbf)&=&\sqrt{\frac{1}{2}\left(1-
    \frac{\epsilon(j,\kbf)}{E(j,\kbf)}\right)}
  .
\end{eqnarray}

In the expression Eq.~(\ref{eq:Cd}) of the charge-charge correlation
function in the sharp resonance approximation, $\alpha$ labels the
two FWS ladders, and not all values of $m,\,m',\,p,\,p'$ ranging from
$-\infty$ to $+\infty$ contribute to a physically observable resonance
in $C_d(\Omega)$. The ``wave-functions'' $\Psi_{m,w}({E}_{\alpha})$
($w=u,v$) have indeed finite extent as a function of $m$, due to
localization on the FWS ladders plotted as a function of the number of
transmitted Cooper pairs [see Fig.~\ref{fig:ladders} (c)]. Further
investigations on the extent of the FWS ladders as a function of $N$
in the semiclassical limit will be presented elsewhere.

\begin{table}[!htb]
  {
     \begin{tabular}{|*{2}{>{\centering\arraybackslash}p{.21\columnwidth}|}*{4}{>{\centering\arraybackslash}p{.11\columnwidth}|}}
     \hline
     {Panel}& {Number of} & $\Gamma_a/\Delta$ &
     $\Gamma_b/\Delta$ & $\Gamma_c/\Delta$ & $\varphi_q/2\pi$\\
     {label} & {terminals} & & & & \\
     \hline \hline
     (a) & 2 & 0.3 & 0.3 & 0 & { } \\
     \hline
     (b) & 2 & 0.4 & 0.2 & 0 & { } \\
     \hline
     (c) & 3 & 0.3 & 0.3 & 0.3 & 0 \\
     \hline
     (d) & 3 & 0.3 & 0.3 & 0.3 & 0.1\\
     \hline \hline
     \end{tabular}
   \caption{The couplings~(a)-(d) used in the numerical
     calculations. [The same labeling~(a)-(d) is used in
       Figs.~\ref{fig:FWS}, \ref{fig:S-Omega}, \ref{fig:log10-S-Omega}
       and~\ref{fig:colorplots-noise}]. The notation $\Gamma_i$ (with
     $i=a,\,b,\,c$) stands for $\Gamma_i=J_i^2/W$, where $J_i$ is the
     hopping amplitude between the dot and lead $S_i$, and $W$ is the
     band-width of the superconductors.
     \label{tab2}}}
\end{table}

\begin{figure*}[htb]
\includegraphics[width=.99\textwidth]{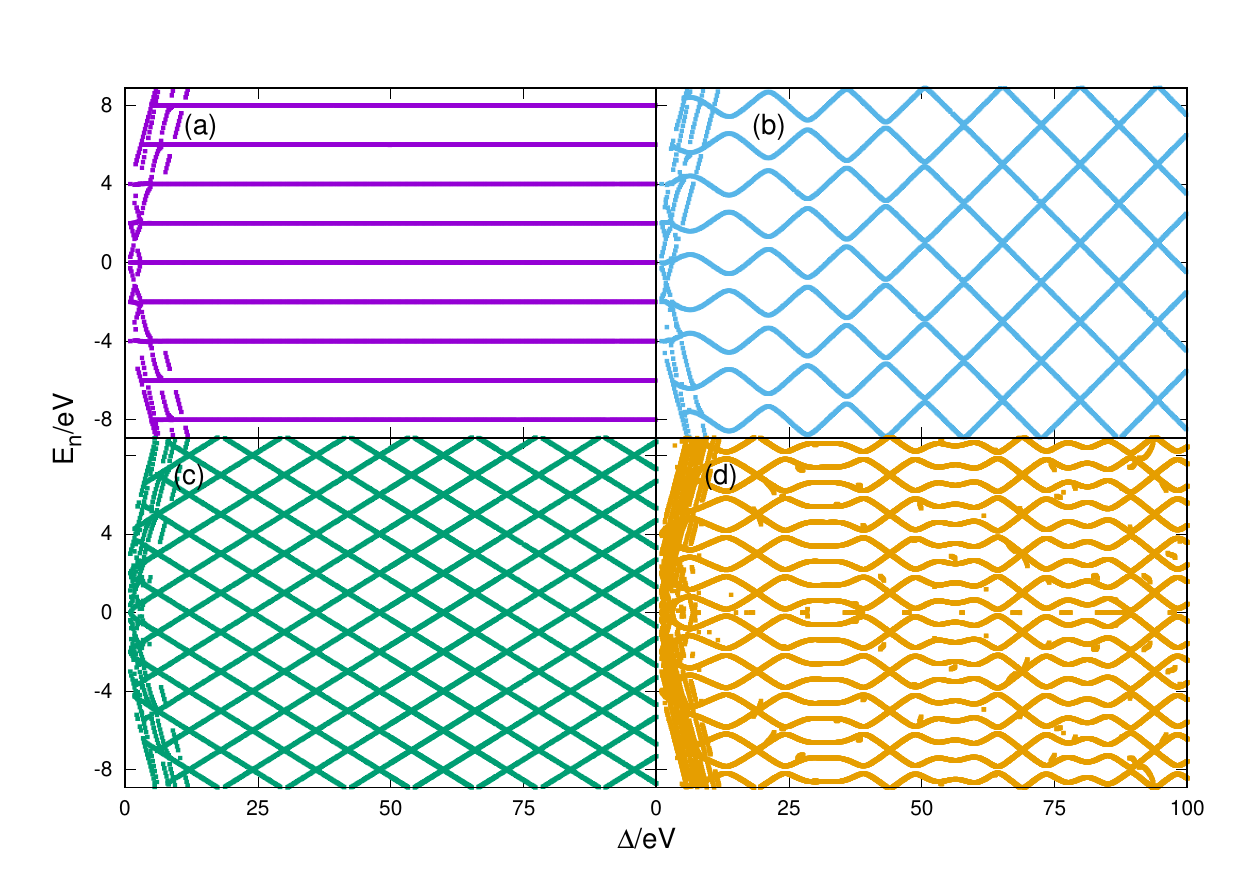}
  \caption{{\it Floquet-Wannier-Stark ladders:} The figure shows the
    Floquet energies as a function of inverse voltage, using the
    rescaled variables given by Eq.~(\ref{eq:energy-shiftter}): The
    $x$-axis is $\Delta/eV$ and $y$-axis is $E/eV$. Small voltage
    means large value of $\Delta/eV$ on $x$-axis. The
    superconducting-QD junction parameters are given in
    Table~\ref{tab2}: (a) Two symmetrically coupled terminals,
    $\Gamma_a/\Delta = \Gamma_b/\Delta = 0.3$; (b) Two terminals with
    generically different couplings to the leads, $\Gamma_a/\Delta =
    0.4,\,\Gamma_b/\Delta = 0.2$; (c) Three symmetrically coupled
    terminals with vanishingly small quartet phase $\varphi_q/2\pi=0$,
    $\Gamma_a/\Delta = \Gamma_b/\Delta = \Gamma_c/\Delta = 0.3$; and
    (d) Three symmetrically coupled terminals with finite value for
    the quartet phase $\varphi_q/2\pi=0.1$, $\Gamma_a/\Delta =
    \Gamma_b/\Delta = \Gamma_c/\Delta = 0.3$. As discussed in the
    text, the spurious data-points for $\Delta/eV\alt 10$ are an
    artifact of the gap edge singularities.
    \label{fig:FWS}
  }
\end{figure*}

\begin{figure*}[htb]
  \includegraphics[width=.99\textwidth]{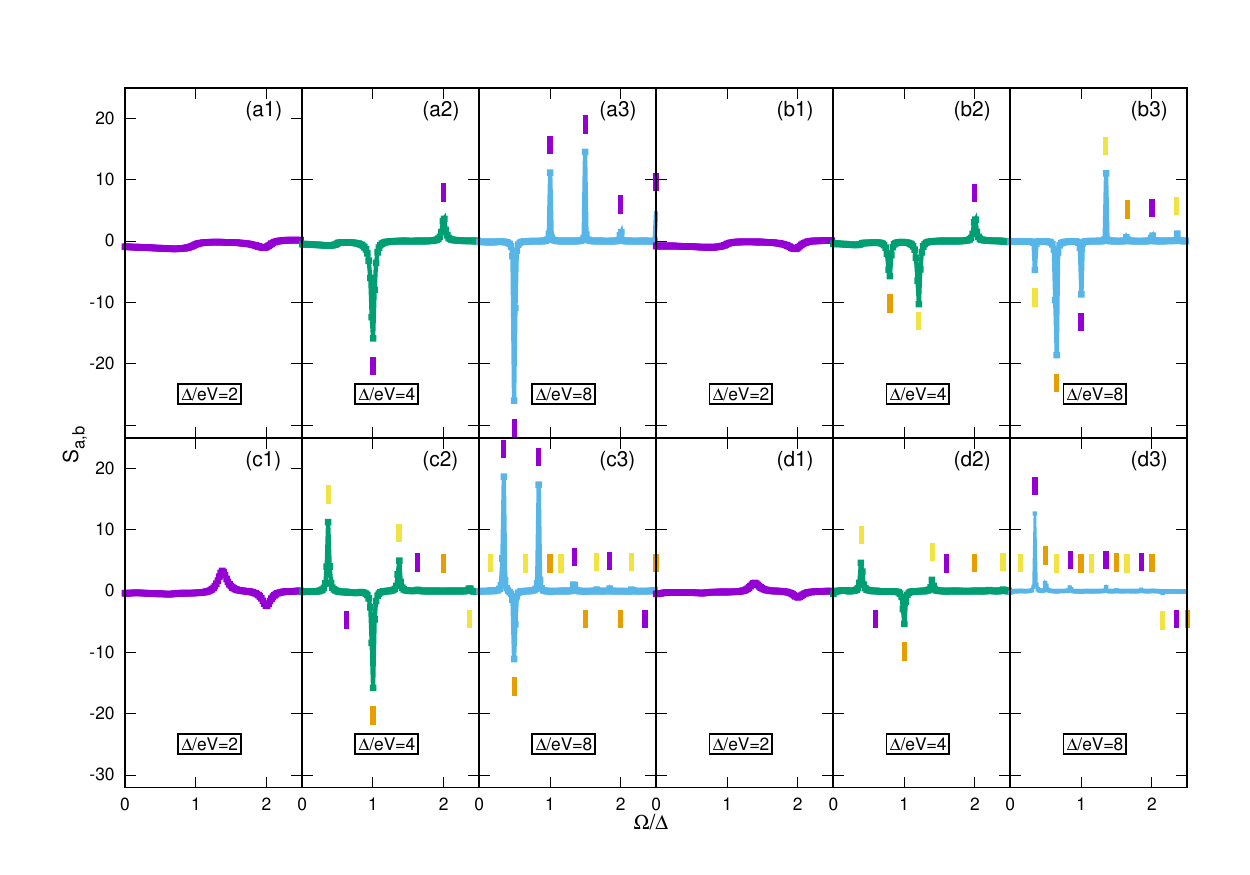}
  \caption{{\it Cross-correlation spectra $S_{a,b}(\Omega)$:} The
    contact transparencies are given in Table~\ref{tab2}: (a1), (a2)
    and (a3) Two symmetrically coupled terminals, $\Gamma_a/\Delta =
    \Gamma_b/\Delta = 0.3$; (b1), (b2) and (b3) Two terminals with
    generically different couplings to the leads, $\Gamma_a/\Delta =
    0.4,\,\Gamma_b/\Delta = 0.2$; (c1), (c2) and (c3) Three
    symmetrically coupled terminals with vanishingly small quartet
    phase $\varphi_q/2\pi=0$, $\Gamma_a/\Delta = \Gamma_b/\Delta =
    \Gamma_c/\Delta = 0.3$; and (d1), (d2) and (d3) Three
    symmetrically coupled terminals with finite value for the quartet
    phase $\varphi_q/2\pi=0.1$, $\Gamma_a/\Delta = \Gamma_b/\Delta =
    \Gamma_c/\Delta = 0.3$.  The values of inverse voltage are
    indicated on the figure: (a1), (b1), (c1) and (d1) $\Delta/eV=2$;
    (a2), (b2), (c2) and (d2) $\Delta/eV=4$; and (a3), (b3), (c3) and
    (d3) $\Delta/eV=8$.  The theoretical prediction [see
      Eqs.~(\ref{eq:E+})-(\ref{eq:E-})] for the collection of values
    of $E_n-E_m$ ({\it i.e.} the differences between Floquet energies)
    is shown by colored bars on each panel. Each bar $x$-axis
    coordinate is at the value of $E_n-E_m$. The color-code is the
    following: yellow, magenta and orange correspond to $\Delta
    E_{+,p}$, $\Delta E_{0,p}$ and $\Delta E_{-,p}$ respectively [see
      Eqs.~(\ref{eq:E+})-(\ref{eq:E-})]. The $x$-axis is
    $\Omega/\Delta$ and $y$-axis is $S_{a,b}$ in natural
    units. {Temperature is vanishingly small.}
    \label{fig:S-Omega}
  }
\end{figure*}

\begin{figure*}[htb]
  \includegraphics[width=.99\textwidth]{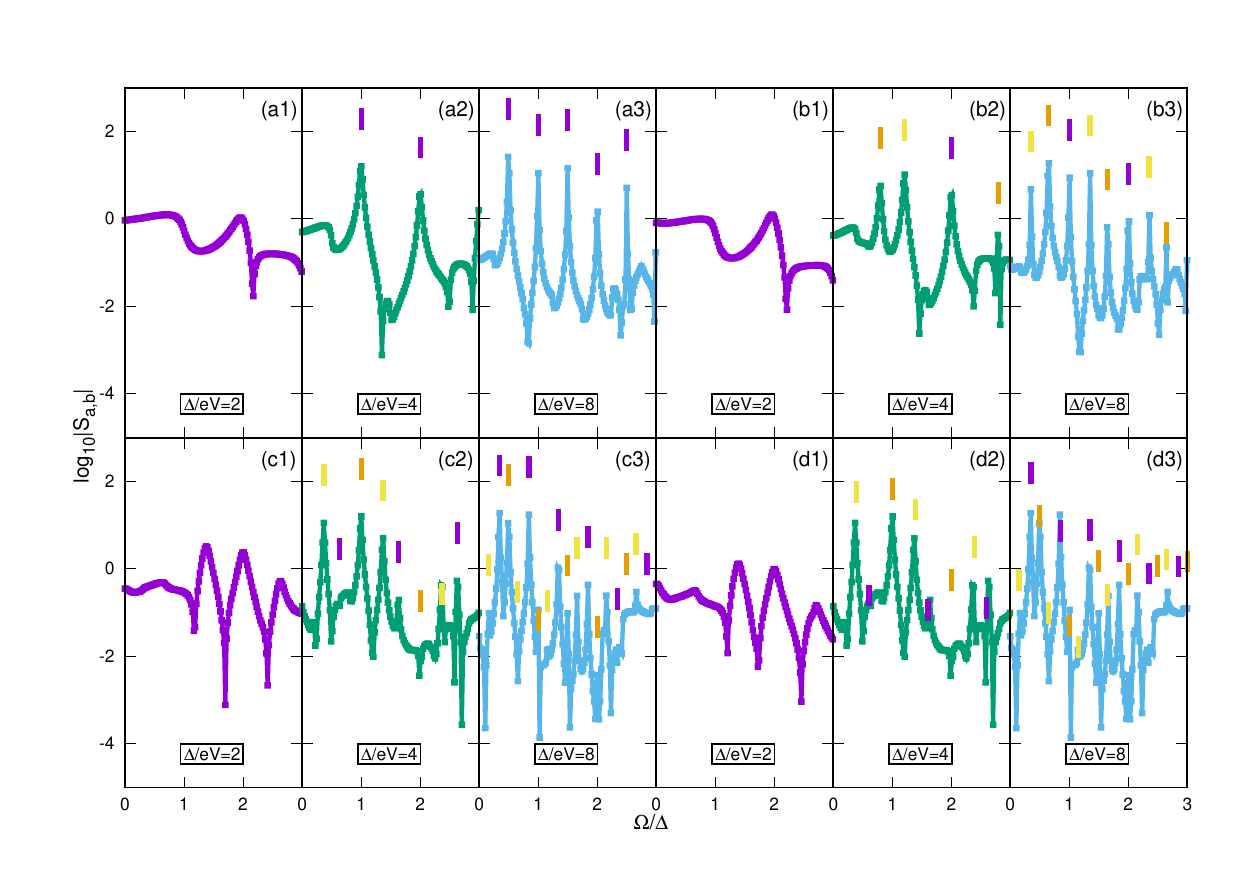}
  \caption{{\it Cross-correlation spectra
      $\log_{10}|S_{a,b}(\Omega)|$:} The same as
    Fig.~\ref{fig:S-Omega} but now $y$-axis ({\it i.e.} the
    $S_{a,b}$-axis) is in log-scale. {Temperature is
      vanishingly small.}
    \label{fig:log10-S-Omega}
  }
\end{figure*}

\begin{figure*}[htb]
    \includegraphics[width=.99\textwidth]{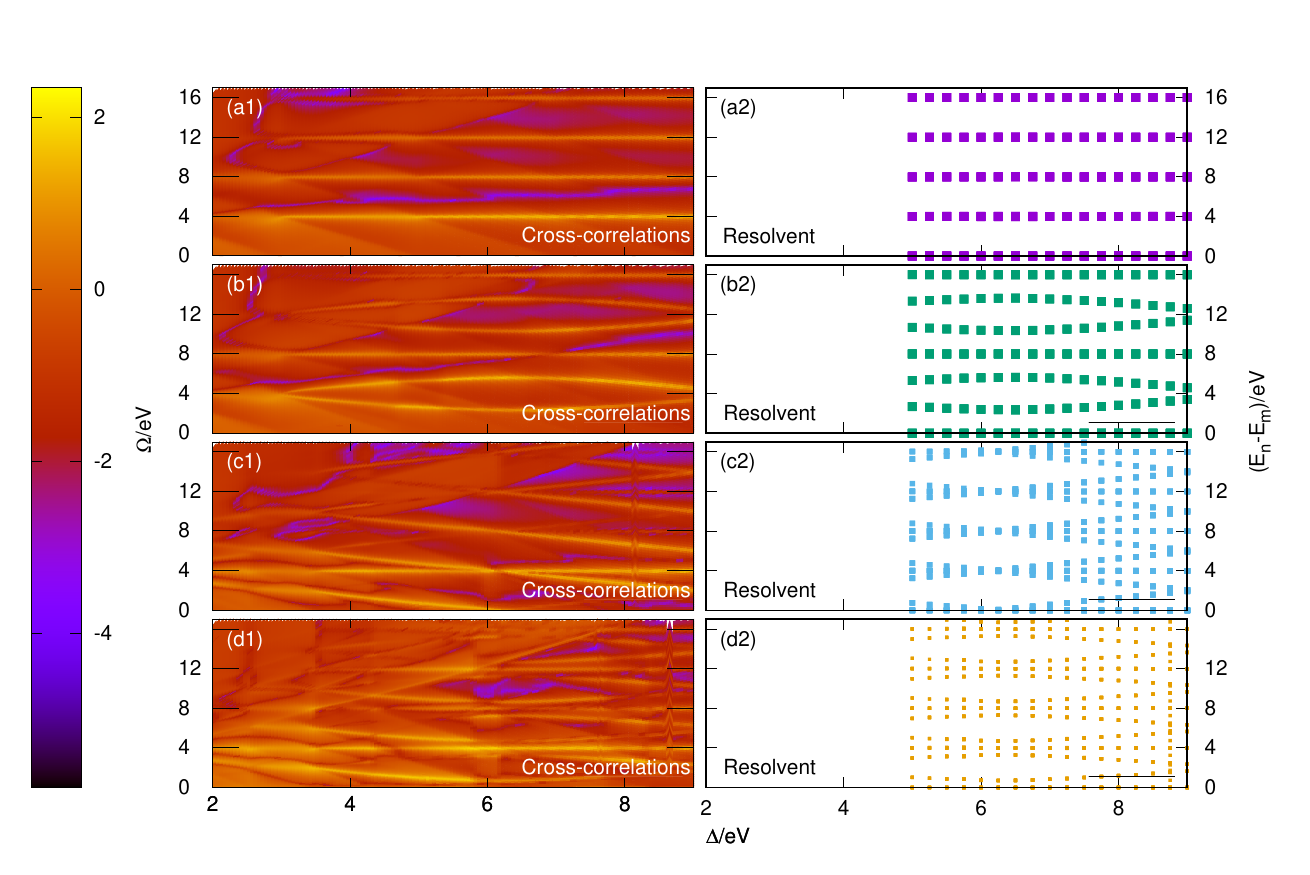}
  \caption{{\it Comparison between the cross-correlation and resolvent
      spectra: } The contact transparencies are given in
    Table~\ref{tab2}: (a1) and (a2) Two symmetrically coupled
    terminals, $\Gamma_a/\Delta = \Gamma_b/\Delta = 0.3$; (b1) and
    (b2) Two terminals with generically different couplings to the
    leads, $\Gamma_a/\Delta = 0.4,\,\Gamma_b/\Delta = 0.2$; (c1) and
    (c2) Three symmetrically coupled terminals with vanishingly small
    quartet phase $\varphi_q/2\pi=0$, $\Gamma_a/\Delta =
    \Gamma_b/\Delta = \Gamma_c/\Delta = 0.3$; and (d1) and (d2) Three
    symmetrically coupled terminals with finite value for the quartet
    phase $\varphi_q/2\pi=0.1$, $\Gamma_a/\Delta = \Gamma_b/\Delta =
    \Gamma_c/\Delta = 0.3$.  Panels [(a1), (b1), (c1), (d1)] show
    $\log_{10}|S_{a,b}(eV/\Delta,\Omega/\Delta)|$ (the logarithm of
    the cross-correlations) in color-scale, as a function of
    normalized inverse voltage $\Delta/eV$ ($x$-axis) and normalized
    frequency $\Omega/eV$ ($y$-axis). Panels [(a2), (b2), (c2), (d2)]
    show the theoretical prediction [see
      Eqs.~(\ref{eq:E+})-(\ref{eq:E-})], {\it i. e.} the collection of
    $(E_n-E_m)/eV$ {\it vs.}  $\Delta/eV$, where $E_n$ and $E_m$ are
    two arbitrary Floquet energies.  The bright peaks in the
    color-plot on panels (a1)-(d1) correspond to the peaks in the
    cross-correlations, also visible in Figs.~\ref{fig:S-Omega}
    and~\ref{fig:log10-S-Omega}. Similar inverse-voltage dependence is
    obtained for the calculated cross-correlations [see panels
      (a1)-(d1)] and the Floquet spectrum ({\it i.e.} the collection
    of $(E_n-E_m)/eV$) [see panels (a2)-(d2)]. The dark blue areas on
    panels~(a1)-(d1) correspond to the regions in which the
    cross-correlations $S_{a,b}$ is small in absolute value (thus with
    negative $\log_{10}|S_{a,b}|$), due to change of sign of $S_{a,b}$
    as a function of $\Omega$ between some of the resonance peaks (see
    Fig.~\ref{fig:S-Omega}). {Temperature is
      vanishingly small.}
    \label{fig:colorplots-noise}
  }
\end{figure*}

In general, Eq.~(\ref{main_approximation_for_R}) is the sum of two
terms because of the summation over $\alpha=\pm$, but one of these can
be discarded if the energy $\tilde{E}\approx E_\alpha$ is close to
$E_\alpha$. Then, the resolvent is well approximated by the resonant
term in Eq.~(\ref{main_approximation_for_R}), which takes a simple
factorized form involving products of the $\Psi$ and $\Phi$
wave-functions. The resulting expression Eq.~(\ref{eq:Cd}) of the
charge-charge correlation function is also factorized into the
corresponding $\Phi$ and $\Psi$-contributions: i) the $S_{i,\alpha}$
(and $S_{j,\beta}$) factors given by Eq.~(\ref{def_S_i_alpha}) depend
only on $\Phi$, and they encode the contributions of the lead $i$ (or
$j$) populations to the resonance $\alpha$ (or $\beta$); ii) The
$\Psi$ terms in the ``form factors'' $F_{\alpha,\beta}$ originate
directly from Wick theorem for products of four-fermion operators, and
they do not depend on the populations in the reservoirs.

Another appealing feature of Eq.~(\ref{eq:Cd}) is that the frequency
dependence solely encoded in a ``minimal'' information about i) the
spectrum of resonances (such as the energies ${E}_{\alpha}$, the
widths $\Gamma_{\alpha}$), and ii) the wave-function
$\Psi_m(E_\alpha)$ at resonance. The $S_{i,\alpha}$ coefficients
contain indeed all information about the stationary state, and they
reflect the initial state of the reservoirs before adiabatic switching
of tunneling processes.

Further semi-classical analysis reveals that the sum over $p$ in
Eq.~(\ref{def_S_i_alpha}) converges easily since the decay of
$\Phi(\tilde{E})_{p}$ at large $p$ is very fast. We note that, besides
this large $p$ behavior, the factor $\nu(j,\alpha,p,\tau)$ diverges
when ${E}_{\alpha}+p\omega_{0}$ is close to $|\Delta_{j}|$. The sum
in~Eq.~(\ref{def_S_i_alpha}) is then dominated by the contribution of
quasiparticle states injected on the dot at energies close to the BCS
gaps in the reservoirs.

A direct consequence of the compact Eq.~(\ref{eq:Cd}) is emergence of
three series of sharp peaks in $S_{a,b}(\Omega)$ at frequencies
\begin{equation}
  \Omega_{\alpha,\beta,p}={E}_\alpha+{E}_\beta+p\omega_0
  \label{eq:freq}
  ,
\end{equation}
which coincide with the preceding Eqs.(\ref{eq:E+})-(\ref{eq:E-}). The
sign of these peaks depends on the Floquet wave-functions, and thus,
it cannot be predicted from simple arguments.

Now, we want to confirm these predictions from independent microscopic
Keldysh Green's functions calculations for the finite frequency
current-current correlation function. We also want to visualize the
frequency-$\Omega$ and voltage-$eV$ dependences of the
cross-correlations, with a choice of the model parameters compatible
with possible experimental realization.

\subsection{Numerical results for the cross-correlation spectra}
\label{sub_current_noise}
In this section, we present our numerical results on the connection
between the FWS ladders of resonances and the symmetrized finite
frequency current-current cross-correlations $S_{a,b}(\Omega)$. After
the necessary definition of $S_{a,b}(\Omega)$, the Floquet spectra are
presented for the four sets of device parameters which will be used
afterwards in the evaluation of $S_{a,b}(\Omega)$.

\paragraph*{Expression of $S_{a,b}(\Omega)$: }
The quantity $S_{a,b}(\Omega)$ calculated numerically is the diagonal
term (in frequency) of the Fourier transform of the following two-time
current-current correlation function
\begin{equation}
  \label{eq:Sab-def}
  S_{a,b}(t,t')=\langle \delta \hat{I}_a(t)\delta \hat{I}_b(t')\rangle
  +\left(t \leftrightarrow t'\right)
  ,
\end{equation}
where $\hat{I}_a(t)$ and $\hat{I}_b(t')$ are the operators for the
currents entering superconducting leads $S_a$ and $S_b$ at times $t$
and $t'$. The notations $\delta \hat{I}_a(t)=\hat{I}_a(t)-\langle
\hat{I}_a(t) \rangle$ and $\delta \hat{I}_b(t')=\hat{I}_b(t')-\langle
\hat{I}_b(t') \rangle$ are used for the deviations with respect to the
expectation value in the stationary states.

\paragraph*{The four sets of device parameters: }
Different configurations of the superconducting-QD ({\it e.g.}
depending on the number of terminals, the symmetry of the contacts and
the presence/absence of quartet phase \cite{Freyn1} for three
terminals) lead to qualitatively different variations in the voltage
dependence of the Floquet spectra.  All numerical calculations
presented below were indeed carried out with the four sets of
parameters labeled by~(a)-(d) in Table~\ref{tab2}. The notation
$\varphi_q$ in this Table stands for the so-called ``quartet phase''
\cite{Freyn1}, corresponding to the static phase combination appearing
in a three-terminal Josephson junction biased with commensurate
voltages. Indeed, the superconducting phase $\varphi_i(t)$ of leads
$S_i$ (with $i=a,b,c$) is given by $\varphi_i(t)=\varphi_i+2 e V_i
t/\hbar$. With opposite voltage biasing in the three-terminal
configurations (c) and (d) ({\it i.e.}  $V_a=-V_b=V$ and $V_c=0$), the
static combination \cite{Freyn1} $\varphi_q$ is given by
$\varphi_q=\varphi_a + \varphi_b-2 \varphi_c\equiv
\varphi_a(t)+\varphi_b(t)-2\varphi_c(t)$.

The rescaled spectra are shown in Fig.~\ref{fig:FWS}, where the
$x$-axis is $\Delta/eV$ (inverse voltage normalized to the gap) and
$y$-axis is $E_n/eV$ (the FWS resonance energies divided by
voltage). Indeed, Eqs.~(\ref{eq:E+})-(\ref{eq:E-}) for the Floquet
levels lead to
\begin{equation}
  \label{eq:energy-shiftter}
  \frac{{E}_{q,\pm}}{eV}=\frac{E_{\pm}}{eV}+q
  .
\end{equation}
These plots in reduced variables \cite{Bentosella} ({\it i.e.} $E/eV$
{\it vs.}  $\Delta/eV$) can advantageously be used instead of the more
conventional ones ({\it i.e.} $E/\Delta$ {\it vs.} $V/\Delta$) in
order to produce regular patterns of avoided levels at low voltage.

Fig.~\ref{fig:FWS} shows the FWS resonance energies $E_n/eV$ evaluated
from the maxima in $|{\cal R}(E)|$ [see Eq.~(\ref{eq:R})]. In addition
to the expected FWS resonances, spurious maxima at energies $E_m^*$
are visible in the small-$\Delta/eV$ region of the spectra. Inspection
of the numerical data for the energy dependence of $|{\cal R}(E)|$
shows that they appear in the vicinity of the gap edges, generally
with tiny curvature $|d^2 |{\cal R}(E_m^*)|/dE^2|$ compared to the
sharp FWS resonances.

Very different behavior emerges on panels (a)-(d) of
Fig.~\ref{fig:FWS}, according to the symmetry of the coupling between
the dot and the leads (see Table~\ref{tab2}). For instance, the basic
period on Fig.~\ref{fig:FWS}~(b) with two terminals is $4eV/\hbar$,
instead of $2eV/\hbar$ for three terminals [see Fig.~\ref{fig:FWS}~(c)
  and Fig.~\ref{fig:FWS}~(d)]. In this case, the Bogoliubov-de Gennes
equations (see section~I in the Supplemental
Material\cite{supplemental}) decouple into two blocks. Each of these
blocks gives rise to a pair of FWS ladders, with the basic period
$\Delta E=4eV/\hbar$. This explains the period doubling observed in
panels (a) and (b) of Fig.~\ref{fig:FWS}.

Another characteristic feature of these spectra of resonances is
absence of level repulsion in Fig.~\ref{fig:FWS}~(a) and
Fig.~\ref{fig:FWS}~(c). For these highly symmetric configurations of
the tunnel amplitudes between the quantum dot and the superconducting
leads, the Bogoliubov-de Gennes Hamiltonian commutes with the
$\sigma^{x}$ Pauli matrix, thus we get two decoupled tight-binding
problems in the Floquet coordinate ({\it e.g.}  the coordinate $N$ on
the $x$-axis of Fig.~\ref{fig:ladders}). This explains why we do not
observe Landau-Zener transitions \cite{Landau,Shevchenko} because the
two FWS ladder are independent in Fig.~\ref{fig:FWS}~(a) and
Fig.~\ref{fig:FWS}~(c). In this case, Bohr-Sommerfeld quantization
\cite{Melin3} becomes exact for a single band, and $E_\pm=\pm\langle
E_{ABS}\rangle$, where $E_{ABS}$ is the average of the equilibrium ABS
energy over the fast phase variable. We have $\pm \langle E_{ABS}
\rangle=0$ for the parameters on Fig.~\ref{fig:FWS}~(a), in agreement
with the horizontal lines seen on this figure.

The spectra shown in Fig.~\ref{fig:FWS}~(b) and Fig.~\ref{fig:FWS}~(d)
are more complex, since they both exhibit repulsion among FWS
resonances.  For panel~(d), the two tunneling paths for Landau-Zener
transitions [instead of a single one for panel~(b)] produce a
modulation of the level repulsion pattern related to
Landau-Zener-St\"uckelberg interferences. \cite{Shevchenko}.

\paragraph*{Numerical results for the cross-correlation spectra: }
Once Fourier transformed, Eq.~(\ref{eq:Sab-def}) is written as a sum
of terms originating from Wick theorem for products of four creation
or annihilation operators in the current-current correlation
function. Each of these terms is given by products of Keldysh Green's
functions \cite{Melin1,Melin2}. The latter can be expressed in terms
of the resolvent defined by Eq.~(\ref{eq:R}). This is formally similar
to the superconducting quantum point contact \cite{Cuevas3} relevant
to break-junction experiments \cite{Cron}. In our calculations of the
current-current cross-correlations in superconducting-QD, the
numerical method relies on recursive Green's functions in energy,
combined to sparse matrices algorithms, and adaptative integration
over the spectral parameter \cite{Melin1,Melin2}. The numerical value
of the cross-correlations converges towards the exact answer upon
increasing the adjustable level of accuracy.

Fig.~\ref{fig:S-Omega} shows the current-current cross-correlations
$S_{a,b}(\Omega)$ [see Eq.~(\ref{eq:Sab-def})] as a function of
frequency $\Omega$, for the four sets (a)-(d) of junction parameters
in Table~\ref{tab2}.  The following values of inverse-voltage are
used: $\Delta/eV=2$ [panels (a1), (b1), (c1), (d1)], $\Delta/eV=4$
[panels (a2), (b2), (c2), (d2)] and $\Delta/eV=8$ [panels (a3), (b3),
  (c3), (d3)]. In agreement with the preceding
section~\ref{section_reduction}, sharp peaks emerge on
Fig.~\ref{fig:S-Omega} for $S_{a,b}(\Omega)$, which become denser and
narrower \cite{Melin3} as $\Delta/eV$ is increased ({\it e.g.}  as
voltage is reduced).  It was verified that the zero-frequency limit
$S_{a,a}(0)$ of the current-current autocorrelation function is always
positive in these calculations, and that the zero-frequency
cross-correlation $S_{a,b}(0)$ is negative for two terminals. The
peaks in the frequency dependence of the cross-correlations
$S_{a,b}(\Omega)$ [see Fig.~\ref{fig:S-Omega}] show both positive or
negative sign, depending on the values of frequency or bias
voltage. Indeed, inspection of Eqs.~(\ref{eq:Cd})-(\ref{eq:F}) reveals
that the sign and amplitude of the resonances in $S_{a,b}(\Omega)$
cannot be fixed by a simple general rule. Instead, it depends on
complex combinations of the Floquet wave-functions which are
oscillating as a function of the coordinate $N$ in
Fig.~\ref{fig:ladders}.

Fig.~\ref{fig:log10-S-Omega} shows the same data as
Fig.~\ref{fig:S-Omega}, but now the $y$-axis ({\it i.e.} the
  $S_{a,b}$-axis) is in $\log$-scale. This reveals many peaks, which
disappear for frequencies $3 \alt \Omega/\Delta$.

The theoretical prediction for the three families of FWS transition
energies $\Delta E_{+,p}$, $\Delta E_{0,p}$ and $\Delta E_{-,p}$ [see
  Eqs.~(\ref{eq:E+})-(\ref{eq:E-})] are shown as bars of different
colors in Figs.~\ref{fig:S-Omega} and~\ref{fig:log10-S-Omega}. The
values of $E_+$ and $E_-$ in Eq.~(\ref{eq:energy-shift}) are
calculated numerically from the sharp maxima in the resolvent $|{\cal
  R}(E)|$ [see Eq.~(\ref{eq:R})]. For clarity, the values of the
$y$-axis ({\it i.e.} the $S_{a,b}$ or $\log_{10}|S_{a,b}|$-axis)
coordinate of all bars has been shifted by a positive or negative
offset.

The $x$-axis ({\it i.e.} the $\Omega$-axis) coordinate of the bars (in
Figs.~\ref{fig:S-Omega} and~\ref{fig:log10-S-Omega}) compares well
with the location in energy of the sharp maxima in $S_{a,b}(\Omega)$
(see Fig.~\ref{fig:S-Omega}) or $\log_{10}|S_{a,b}(\Omega)|$ (see
Fig.~\ref{fig:log10-S-Omega}). This provides numerical evidence for
the expectation (discussed above in the preceding
section~\ref{section_parallel}) that the peak frequencies match the
energy differences $E_n-E_m$ between pairs of Floquet states [see also
  the preceding Eq.~(\ref{eq:freq}) deduced from
  Eqs.~(\ref{eq:C})-(\ref{eq:F})]. We note that a few of the
theoretically predicted peaks are barely visible in $S_{a,b}(\Omega)$
(see Fig.~\ref{fig:S-Omega}) or $\log_{10}|S_{a,b}(\Omega)|$ (see
Fig.~\ref{fig:log10-S-Omega}), because they are directly surrounded by
sharp peaks with positive and negative signs, and thus, the value
$|S_{a,b}|$ is weak for these resonances.

Panels (a1)-(d1) of Fig.~\ref{fig:colorplots-noise} show the
cross-correlations $S_{a,b}(\Delta/eV,\Omega/eV)$ in the plane of
parameters $\Delta/eV$ (on $x$-axis) and $\Omega/eV$ (on
$y$-axis). This is compared with panels (a2)-(d2) on the same figure,
featuring $(E_n-E_m)/eV$ (on $y$-axis) as a function of inverse
voltage $\Delta/eV$ (on $x$-axis). The data for the resolvent are
similar to those in Fig.~\ref{fig:FWS}, but now in the experimentally
relevant window $2\alt \Delta/eV\alt 10$ of inverse voltage. The data
on panels (a2)-(d2) were truncated to $\Delta/eV \ge 5$. The values of
$E_n$ and $E_m$ calculated from $|{\cal R}(E)|$ are indeed within the
gap region $-\Delta\alt E_n,\,E_m \alt \Delta$. This implies lower
bound on $\Delta/eV$ if one wants to produce from $|{\cal R}(E)|$ the
full spectrum of $(E_n-E_m)/eV$ within range $0<(E_n-E_m)/eV<16$ [see
  panels~(a2)-(d2) of Fig.~\ref{fig:colorplots-noise}].

We note also the presence in Fig.~\ref{fig:colorplots-noise}~(d1) of a
resonance line at frequency $\Omega=2\Delta$, in addition to the FWS
resonance lines coinciding with
Fig.~\ref{fig:colorplots-noise}~(d2). This resonance corresponds to
the expected transitions between both gap edge singularities at
energies $\pm \Delta$. The resulting peak in $S_{a,b}$ cannot be
distinguished from the series of transitions between FWS ladders in
the preceding Figs.~\ref{fig:S-Omega} and~\ref{fig:log10-S-Omega},
because $\Delta/eV$ is an integer on these figures, implying that the
quasiparticle resonance at energy $2\Delta$ necessarily coincides with
the transitions $\Delta E_{0,p}$ [see Eq.~(\ref{eq:DeltaE0p})].

It is concluded from Figs.~\ref{fig:S-Omega},~\ref{fig:log10-S-Omega}
and~\ref{fig:colorplots-noise} that the frequency $\Omega$ of the
sharp peaks in the cross-correlation matches very well the energy
difference $E_n-E_m$ between FWS resonances in the resolvent, in
addition to a quasiparticle line at energy $\Omega=2\Delta$.  This
confirms that finite frequency cross-correlations can be used to make
spectroscopy of the Floquet spectrum ({\it i.e.}  the spectrum of the
FWS ladders) in a multiterminal superconducting-QD. Our predictions
for the voltage dependence is of particular relevance to experiments,
which is discussed now in the concluding section.

\section{Summary and perspectives}
\label{Conclusion}

\subsection{Summary}
The equilibrium ABS are protected by a finite energy gap from the
semi-infinite quasiparticle continua in multiterminal
superconducting-QD. The nonequilibrium Floquet states can thus have
very long life-time. Still, the higher-order MAR provide a finite
width $\gamma$ to these FWS resonances, which is exponentially small
in $\Delta/eV$. Due to the drastic reduction of $\gamma$ as
$eV/\Delta$ decreases, sharp resonances emerge already in a relatively
large range of $eV/\Delta \alt 1/3$ (see
Figs.~\ref{fig:S-Omega} and~\ref{fig:log10-S-Omega} for the evolution
with voltage of the width of the resonances in the
cross-correlations).

An analytical theory was presented, which takes advantage of the
smallness of the width of the FWS resonances. A compact expression was
obtained for the charge-charge correlation function. Remarkably, due
to the presence of sharp FWS resonances, the charge correlation
function factorizes into a product of two quantities: i) a first term
containing information about the populations in the lead, and ii) a
second one involving products of four Floquet wave-functions on the
QD. This analytical theory reveals three series of peaks in the
frequency dependence of the cross-correlations, which receive
interpretation of transitions between Floquet states belonging to the
same or different ladder. Numerical calculations for the
cross-correlations $S_{a,b}$ (between currents entering the
superconducting leads $S_a$ and $S_b$) were presented in the case of
two- and three-terminal configurations with various symmetries of the
couplings. The numerical results for the location in energy of the
resonances are in a quantitative agreement with the analytical theory.

It is concluded that the nontrivial voltage dependence of the Floquet
spectrum can be accessed experimentally {\it via} finite frequency
noise spectroscopy in a superconducting-QD.

\subsection{Perspectives}

{Calculating the nonsymmetrized correlators instead of
  the symmetrized ones would not change the energy values of the peaks
  in the noise, but may possibly modify their sign. This question will
  be addressed in the future in connection with experiments.}

Expanding the modulus square in Eq.~(\ref{def_S_i_alpha}) produces
several terms in which different physical processes can be
recognized. It would be interesting to investigate a similar expansion
for the current cross-correlations.  Classifying the different
processes beyond perturbation theory was already done for the
dc-current \cite{Melin4} and for the current cross-correlations
\cite{Melin5,Golubev1,Freyn2,Floser,Golubev2} of a metallic normal
metal-superconducting-normal metal double junction. For a
multiterminal all-superconducting-QD (see
Fig.~\ref{fig:schema-N-terminaux}), this can lead to nonperturbative
characterization of the quartet \cite{Freyn1,Pfeffer,Cohen}, multipair
or phase-MAR\cite{Jonckheere,Cohen,Note1} channels. At present time,
separation of the dc-current \cite{Jonckheere} and the current-current
cross-correlations \cite{Melin2} into the different physical channels
relies solely on the symmetries with respect to phase or voltage
inversion \cite{Melin6}.

Another interesting perspective is to develop nonstandard algorithms
to calculate the current and the current-current correlation function
in the sharp resonance approximation. Namely, this approximation is
based on a limited number of parameters for the spectrum of FWS
resonances ({\it i.e.} the position of the Floquet resonance energies,
and their width), and on the Floquet wave-functions at resonance. The
factorized form of the charge-charge correlation function [see
  Eq.~(\ref{eq:Cd})] suggests the possibility of spectacular
enhancement of the performances of the codes with respect to those used
in Sec.~\ref{sub_current_noise} to evaluate the current correlation
functions.

It is also an open question to show that FWS ladders are robust
against Coulomb interaction. We note that finite frequency noise has
been calculated recently for an interacting quantum dot
\cite{Crepieux}.

Fig.~\ref{fig:colorplots-noise}~(b1) and
Fig.~\ref{fig:colorplots-noise}~(b2) reveal nontrivial features
already for a two-terminal device with generically different couplings
to the leads in the experimentally accessible voltage range $0.1 \alt
eV/\Delta \alt 1$.  Two-terminal devices are much easier to control
experimentally than their three-terminal
counterparts. Fig.~\ref{fig:colorplots-noise}~(b1) and
Fig.~\ref{fig:colorplots-noise}~(b2) feature repulsion among FWS
resonances in this case, which is a signature of quantum coherent
coupling between the time-periodic states originating from different
ladders. The Floquet wave-functions are then delocalized on both ``FWS
ladder rungs'' in quasicoincidence (as a function of the coordinate
$N$ in Fig.~\ref{fig:ladders}). Future perspectives on a ``Floquet
qu-bit'' based on the time-periodic dynamics of the superconducting-QD
can be envisioned, including opening on the physics of driven
qu-bits\cite{Leek,Murch}.

{Finally, the possibility to perform tunnel
  spectroscopy of the FWS ladders is currently under investigation.}

\acknowledgements{The authors thank the Centre R\'egional Informatique et
d'Applications Num\'eriques de Normandie (CRIANN) for the use of its
facilities.  R.M. and R.D. acknowledge financial support from the
Centre National de la Recherche Scientifique (CNRS) and Karlsruhe
Institute of Technology (KIT), through the International Laboratory
``LIA SUPRADEVMAT'' between the Grenoble and Karlsruhe campuses. This work was 
partly supported by Helmholtz society through program STN and the DFG via the projects DA 1280/3-1.}



\end{document}